\documentclass[12pt]{article}
\usepackage{epsf,graphics}

\renewcommand{\thefootnote}{\#\arabic{footnote}}
\newcounter{counterA}

\begin{document}

\newcommand{\gtrsim}{ \mathop{}_{\textstyle \sim}^{\textstyle >} }
\newcommand{\lesssim}{ \mathop{}_{\textstyle \sim}^{\textstyle <} }

\renewcommand{\thefootnote}{\fnsymbol{footnote}}
\setcounter{footnote}{0}
\begin{titlepage}

\def\thefootnote{\fnsymbol{footnote}}

\begin{center}

\hfill TU-676\\
\hfill hep-ph/0212002\\
\hfill November, 2002\\
\vskip .5in

{\Large \bf
Recent Muon $g-2$ Result \\
in Deflected Anomaly-Mediated Supersymmetry Breaking
}

\vskip .45in

{\large
Nobutaka Abe and  Motoi Endo
}

\vskip .45in

{\em
Department of Physics,  Tohoku University\\
Sendai 980-8578, Japan
}

\end{center}

\vskip .4in

\begin{abstract}
We study the deflected anomaly-mediated supersymmetry breaking (AMSB) 
scenario in the light of the recent result of the muon $g-2$ from 
Brookhaven E821 experiment.  The E821 result suggests the deviation 
from the SM prediction, though there remain unsettled uncertainties.  
We find that the supersymmetric contribution to the muon $g-2$ can 
be $\mathcal{O}(10^{-9})$, large enough to fill the deviation, 
with other experimental constraints satisfied.  
In particular, the Higgs mass and $b \to s \gamma$ put severe 
constraints on the model and large $\tan\beta$ is favored 
to enhance the muon $g-2$.  
\end{abstract}
\end{titlepage}

\renewcommand{\thepage}{\arabic{page}}
\setcounter{page}{1}
\renewcommand{\thefootnote}{\#\arabic{footnote}}
\setcounter{footnote}{0}

\setcounter{equation}{0}

Recently, the new muon $g-2$ result has been announced from 
the Brookhaven E821 experiment~\cite{BNL2002},
\begin{eqnarray}
  a_\mu({\rm E821}) = 11\ 659\ 203(8) \times 10^{-10}\ ,
\end{eqnarray}
where the error has become comparable to that of the standard model (SM) 
prediction.  The SM theoretical value of the muon $g-2$ has been 
reported in Refs.~\cite{Czar1999}--\cite{Knecht2001}.  
The main sources of the uncertainties of the SM prediction come from 
the leading hadronic vacuum polarization and the light-by-light 
contributions.  As for the leading hadronic contribution, we have to rely 
on the experimental data, that is, $e^+e^-$ cross section and hadronic 
$\tau$ decay data, where the $\tau$ decay data is translated into 
$e^+e^-$ cross section by assuming the isospin symmetry.  
The most recent evaluations are given in Table~\ref{table:hadronic}.  
In the new detailed evaluation by Davier {\it et al}~\cite{Davier2002}, 
they carefully considered radiative corrections to the $e^+e^-$ 
cross section and took into account the isospin symmetry breaking effects 
explicitly.  From Table~\ref{table:hadronic}, the results of 
$e^+e^-$ and $\tau$-based are, unfortunately, inconsistent and the 
origin of this difference has not been clarified.  As for the $e^+e^-$-based 
evaluation,  we notice that the uncertainty becomes comparable to the 
previous works which uses the $\tau$ decay data, and the independent 
analysis by Hagiwara {\it et al}~\cite{Hagiwara2002} gave a similar result.  
In this letter, we are inclined to use the $e^+e^-$-based result 
by Davier {\it et al} for the leading hadronic contribution.  
Then, with the corrected sign of the light-by-light 
contribution~\cite{Knecht2001}, the SM prediction becomes 
\begin{eqnarray}
  a_\mu({\rm SM}) &=& 11659169.1(7.8)\times 10^{-10}
  \quad{\rm [e^+e^--based]}\ .
\end{eqnarray}
Therefore the difference between the experimental value and the 
SM prediction is 
\begin{eqnarray}\label{eq:E821}
  a_\mu({\rm E821}) - a_\mu({\rm SM}) &=& 33.9(11.2) \times 10^{-10}
  \quad{\rm [e^+e^--based]}\ ,
\end{eqnarray}
which means that the deviation is 3.0$\sigma$~\footnote{
  The E821 result is consistent with the $\tau$-based prediction 
  at 1.6$\sigma$ level.  
}.  
Though the uncertainties of the SM prediction have not been settled, 
there remains the possibility for the deviation to be physical.  
If this deviation is a signal of new physics, additional contribution 
to the muon $g-2$ is required to be $\mathcal{O}(10^{-9})$.  
\begin{table}[tb]
  \begin{center}
    \begin{tabular}{ccc}
      \hline
      {Authors} & {$a_\mu({\rm had,L.O.}) \times 10^{10}$} & {Data} \\
      \hline
      {Davier and H\"{o}cker~\cite{Davier1998}} & {$692.4(6.2)$} & 
      {$[e^+e^-,\tau]$} \\
      {Narison~\cite{Narison2001}} & {$702.1(7.6)$} & 
      {$[e^+e^-,\tau]$} \\
      {Troc\'{o}niz and Yndur\'{a}in~\cite{Troconiz2002}} & {$695.2(6.4)$} & 
      {$[e^+e^-,\tau]$} \\
      {Davier {\it et al}~\cite{Davier2002}} & {$684.7(7.0)$} & 
      {$[e^+e^--{\rm based}]$} \\
      {Davier {\it et al}~\cite{Davier2002}} & {$701.9(6.2)$} & 
      {$[\tau-{\rm based}]$} \\
      {Hagiwara {\it et al}~\cite{Hagiwara2002}} & {$683.1(6.2)$} &
      {$[e^+e^--{\rm based}]$} \\
      \hline
    \end{tabular}
    \caption{The evaluation of the leading hadronic vacuum polarization 
      contribution}
    \label{table:hadronic}
  \end{center}
\end{table}

Supersymmetry (SUSY) is one of the most motivated models which 
extend the SM and the muon $g-2$ has been investigated in SUSY 
models~\cite{SUSYg-2,MDMonSUSY,recentMDM}.  These models often provide 
the universal gaugino mass.  However once SUSY is extended to include 
the gravity, the quantum effects via the super-Weyl anomaly~\cite{SW-anomaly} 
always manifest themselves in the soft SUSY breaking terms and 
give another class of gaugino mass spectra.  
This SUSY breaking mediation mechanism is known as anomaly-mediated 
SUSY breaking (AMSB)~\cite{AMSB}.  
Though anomaly-mediated effects may be small compared to 
the other SUSY breaking effects, there are some cases where AMSB become 
dominant.  In fact, AMSB dominates if the SUSY breaking sector has no 
direct interactions with the minimal supersymmetric standard model (MSSM) 
sector but gravitation.  In this letter, we study models where 
the anomaly-mediated effects dominate.  
Though AMSB has attractive features~\footnote{
  In AMSB, dangerous CP and flavor violating parameters are naturally 
  suppressed.  The AMSB dominant models may also provide a solution to the 
  gravitino problem and the cosmological moduli problem~\cite{NPB570-455}.  
}, the original model~\cite{RS} suffers from the tachyonic slepton problem.  
Furthermore a parameter set which survives the $b \to s \gamma$ bound 
generally leads a negative contribution to the muon $g-2$.  

Various attempts have been made to avoid the tachyonic 
slepton~\cite{tachyon}.  The deflected AMSB model~\cite{deflected-AM,RSW} 
is a successful and attractive one.  In particular, this model makes 
the sign of the wino mass $M_2$ and the gluino mass $M_3$ {\it identical}.  
This is very important for the recent result of the muon $g-2$.  
In fact, the experimental results of the muon $g-2$ and the branching ratio 
of $b \to s \gamma$ favor both $\mu_H M_2$ and $\mu_H M_3$ positive, 
where $\mu_H$ is Higgs mixing parameter.  On the other hand, the pure AMSB 
model predicts opposite sign of the wino and gluino masses, and some 
additional mechanism is required to modify the gaugino mass relations.  
In fact, the muon $g-2$ has been investigated in the minimal AMSB 
scenario~\cite{mAMSB}, where only the sfermion sector is modified.  
In this scenario, the sign of wino and gluino mass is opposite.  
Thus if we regard the deviation between $a_\mu({\rm E821})$ and 
$a_\mu({\rm SM})$ given in Eq.(\ref{eq:E821}) as a signal of new physics, 
this model conflicts with the recent muon $g-2$ result and 
the $b \to s \gamma$ bound.  Thus the gaugino sector should be modified.  
Such additional effects, however, generally induce new CP violating phases 
in the gaugino masses.  The deflected AMSB scenario is the only known model 
in AMSB which provides a natural solution to both of these problems.  

The deflected AMSB model is safe against CP and flavor violations~\footnote{
  We assume that the CP phase from $B$ parameter is also suppressed.  
  A mechanism for the suppression has been proposed in the context of 
  the deflected AMSB~\cite{deflected-AM}.  
} and provides a preferred sign of a SUSY contribution to the muon $g-2$.  
However, the Higgs mass and the $b \to s \gamma$ branching ratio 
put severe constraints on the parameter space.  
In the deflected AMSB squarks are generally 
not so heavy compared to other SUSY breaking scenarios 
and the Higgs mass bound requires relatively large soft masses.  
Thus the SUSY contribution to the muon $g-2$ becomes rather small.  
As a result, large $\tan\beta$ is required to enhance the muon $g-2$ 
and thus the $b \to s \gamma$ branching ratio bound becomes dangerous.  
Hence the detailed analysis is required for the deflected AMSB whether 
this model is phenomenologically viable when the SUSY contribution to 
the muon $g-2$ is as large as $\mathcal{O}(10^{-9})$.  
An investigation of the muon $g-2$ in the deflected 
AMSB scenario has been performed by Abe {\it et al}~\cite{axion}, 
where hadronic axion model in this framework are studied.  However 
Abe {\it et al} made less general analysis.  In fact, the parameter space 
is specified to provide realistic axion decay constant.  Moreover 
the Higgs mass is also calculated by using the effective potential 
at one loop level.  However the higher order corrections become 
important for the Higgs mass.  Hence we reanalyze the deflected AMSB model 
in more general setting.  

Let us first review the soft masses and some properties of the deflected 
AMSB model~\cite{JHEP9905-013,NPB559-27}.  The anomaly-mediated effects 
on the soft masses can be given by inserting the compensator field, 
$\Phi\equiv 1+F_\Phi\theta^2$.  Here $F_\Phi$ is vacuum expectation
value (VEV) of the scalar auxiliary field in the gravitational 
supermultiplet and takes a value of order of the gravitino mass.  
In order to avoid tachyonic slepton and modify 
gaugino masses, we introduce a singlet field $X$ whose auxiliary 
component has non-zero VEV by the following superpotential.  We also add 
$N_{\bf 5}$-pairs of ${\bf 5}=(Q,L)$ and ${\bf \bar{5}}=(\bar{Q},\bar{L})$ 
of $SU(5)$ to mediate the SUSY breaking from the singlet field $X$ to MSSM 
sector.  Then the additional terms in the superpotential consist of 
the following two parts;
\begin{eqnarray}\label{W_X}
  W_{X{\bf\bar{5}5}} = \lambda_Q X \bar{Q} Q + \lambda_L X \bar{L} L\ ,
\end{eqnarray}
and the non-renormalizable term
\begin{eqnarray}
  W_{X} &=& \frac{ 1 }{ \Lambda^{n-3}\Phi^{n-3} } X^n\ ,
\end{eqnarray}
where $\Lambda$ is a some mass parameter which is assumed to be of order of 
the Plank scale, and $n$ is a positive integer but larger than three. 
By minimizing the scalar potential of $X$, the VEVs of the scalar and 
auxiliary components of $X$ are determined.  In particular, the auxiliary 
component VEV becomes $F_X/\langle X \rangle = F_\Phi(n-3)/(n-1)$ and 
obviously we do not introduce new CP violating phase.  

Once the scalar component of $X$ acquires the VEV, 
$N_{\bf 5}$-pairs of $(Q,L)$ and $(\bar{Q},\bar{L})$ have masses 
$M_{\rm mess} = \lambda\langle X\rangle$ and play the same role 
as the messenger fields in GMSB.  
At the messenger scale $M_{\rm mess}$, the chiral superfields $(Q,L)$ and 
$(\bar{Q},\bar{L})$ decouple and the threshold effects induce the additional 
corrections to the soft parameters.  As a result, the soft parameters 
at the messenger scale are given by
\begin{eqnarray}
  M_{\lambda}(M_{\rm mess}) &=& 
  -\frac{ \alpha_i(M_{\rm mess}) }{ 4\pi } 
  \left[ b_i - \frac{2}{n-1} N_{\bf 5} \right] F_\Phi\ ,
  \label{M_lam}\\
  m^2_{\tilde{f}}(M_{\rm mess}) &=& 
  \frac{1}{(4\pi)^2}
  \left[ 2C_i^f \left(b_i - \frac{4(n-2)}{(n-1)^2} N_{\bf 5} \right)
    \alpha^2_i(M_{\rm mess})
  \right.
  - N_u\alpha_t(M_{\rm mess})
  \left\{
    \frac{13}{15} \alpha_1(M_{\rm mess}) 
  \right.
  \nonumber \\ && 
  \left.
    \left.
      + 3 \alpha_2(M_{\rm mess})
      + \frac{16}{3} \alpha_3(M_{\rm mess})
      - 6\alpha_t(M_{\rm mess})
    \right\} 
  \right] |F_\Phi|^2\ ,
  \label{m^2_f}\\
  A_f(M_{\rm mess}) &=& 
  -\frac{ y_f(M_{\rm mess}) }{ 4\pi } \sum_{fields \in f}
  \left[ 2 C_i^f \alpha_i(M_{\rm mess}) 
    - N_u \alpha_t(M_{\rm mess})
  \right] F_\Phi\ ,
  \label{A_f}
\end{eqnarray}
where $\alpha_t$ is the top-quark Yukawa coupling, $\alpha_i$s are the gauge 
coupling constants ($i$ is the index of the SM gauge groups), $b$ is the 
$\beta$ function $b=(-\frac{33}{5},-1,3)$, and $C^f$ is the second-order 
Casimir.  The parameter $N_u$ takes $N_u=(1,2,3)$ for $\widetilde{q}_L^{3rd},
\widetilde{t}_R$ and $h_u$ and $N_u=0$ for other particles.  Thus the soft 
SUSY breaking masses are of order $m_{\rm SUSY} \sim F_\Phi\alpha/4\pi$.  
Phenomenologically, $F_\Phi$ and the gravitino mass are required to be 
$\mathcal{O}(10^1-10^2{\rm TeV})$.  

There is also the case of not including $W_{X}$ in the superpotential.  
In this case, the soft masses take the identical values to $n=3$ in 
eqs.(\ref{M_lam})--(\ref{A_f}), thus this model is often called 
the case of ``$n = 3$'' and in this letter, we refer to this case as 
``$n = 3$''.  
We note $X$ is a flat direction and only lifted by the pure AMSB effect, 
but the scalar potential of $X$ is not stabilized.  There are some 
approaches to stabilize the potential by introducing a extra UV 
free gauge symmetry~\cite{RSW} or higher dimensional terms in the 
K\"ahler potential~\cite{axion}.  Then the VEV of $X$ can be determined.  
For the sake of the generality of the model, we regard the VEV of $X$ as 
a free parameter in the following.  

We also note here the issue of the lightest superparticle (LSP). 
In the case of $n=3$, the mass of the fermionic part of 
the singlet field, $\tilde{X}$, is approximately
\begin{eqnarray}
  M_{\tilde{X}} \simeq 
  \frac{ N_{\bf 5}\lambda^2 }{ 16\pi^4 }
  \left[ g_3^2 
    - \frac{ 5(5N_{\bf 5}+2) }{ 8 } \lambda^2
  \right] F_\Phi^\dagger\ ,
\end{eqnarray}
where we assume $\lambda \equiv \lambda_Q \simeq \lambda_L\lesssim O(1)$ and  
neglect $U(1)_Y$ and $SU(2)_L$ gauge couplings.  One can see that the mass of 
$\tilde{X}$ arises at the two-loop level and is much lighter than any 
other mass of the superparticle in the MSSM sector.  
Thus $\tilde{X}$ becomes the LSP in this case.  
On the other hand, for $n \geq 4$, the mass of $\tilde{X}$ is as heavy as 
$F_\Phi\sim \mathcal{O}(10^1-10^2{\rm TeV})$.  In this case, 
we investigate which particle will be the LSP in the MSSM sector.  

We summarize the parameters in the model.  There are six parameters: 
\begin{eqnarray}\label{eq:free_parameters}
  F_\Phi,~~M_{\rm mess},~~N_{\bf 5},~~n,~~\tan\beta,~~sign(\mu_H)\ .
\end{eqnarray}
We fix these parameters at the messenger scale $M_{\rm mess}$ and solve one 
loop renormalization group equation from the messenger scale to 
the weak scale.  Then we determine the magnitude of higgsino mass $\mu_H$ 
and Higgs mass parameter $B$ by electroweak symmetry breaking (EWSB) 
conditions with the Higgs potential at one loop order.  

Now we analyze the muon $g-2$ in the model.  
At the weak scale, we calculate the SUSY contribution to the muon $g-2$.  
The SUSY contribution consists of chargino-sneutrino and neutralino-smuon 
diagrams.  With some approximations, the SUSY contribution is expressed 
as~\cite{SUSYg-2}
\begin{eqnarray}
  a_\mu\ ({\rm SUSY}) = 
  \frac{5 \alpha_2}{48\pi} \frac{m_\mu^2}{m_{\rm SUSY}^2} 
  sign(M_2 \mu_H) \tan\beta\ ,
  \label{a_susy}
\end{eqnarray}
where $m_\mu$ is the muon mass and $m_{\rm SUSY}$ is a typical mass of the 
superparticles in the loop diagrams.  From this equation, we find some 
important features.  First, the SUSY contribution to the muon $g-2$ is 
enhanced when $\tan\beta$ is large.  Second, $a_\mu({\rm SUSY})$ decreases 
as $m_{\rm SUSY}$ become heavier.  Moreover, the recent E821 result 
suggests $sign(M_2 \mu_H)$ to be positive.  

Some experiments provide constraints on the model.  In particular, 
the Higgs boson mass bound is severe.  Generally, SUSY models predict not 
so large Higgs mass.  Therefore the heavy stops or the large top trilinear 
coupling $A_{\tilde{t}}$ are required~\cite{SUSY_Higgs} to satisfy the 
lower bound from the LEP II experiment~\cite{HiggsMass}
\begin{eqnarray}
  m_h \geq 114.1\ {\rm GeV}\ .
\end{eqnarray}
Since in the deflected AMSB model, the trilinear coupling is not so large, 
stops should be heavy.  That is, $F_\Phi$ is required to be large.  
In particular, the model tends to predict relatively small gluino mass 
at the messenger scale.  Consequently squarks do not receive 
large renormalization group corrections from gluino.  Thus the Higgs mass 
bound requires the whole superparticles to be relatively heavy and 
hence the muon $g-2$ tends to be suppressed.  In the numerical analysis, 
we use the {\tt FeynHiggsFast} package~\cite{FeynHiggsFast} 
to compute the lightest Higgs boson mass.  

The muon $g-2$ is enhanced by increasing $\tan\beta$.
Then the constraint from the inclusive $b \to s \gamma$ decay becomes 
important.  The experimental measured value of ${\rm Br}(b \to s \gamma)$ 
is consistent with the SM prediction.  On the other hand, SUSY contributions 
may significantly change the SM prediction of ${\rm Br}(b \to s \gamma)$.  
The SUSY contribution to ${\rm Br}(b \to s \gamma)$ mainly consists of 
charged Higgs-top and chargino-stop diagrams and they are enhanced 
for large $\tan\beta$.  Thus when $\tan\beta$ is large, dominant parts of 
the SUSY contribution have to cancel each other.  Such cancellation occurs 
when $sign(M_3 \mu_H)$ is positive and this situation is naturally given 
by the deflected AMSB scenario.  In this letter, we estimate the SM 
contribution according to Ref.~\cite{Kagan:1999ym}.  
As for the charged Higgs contribution, we use the next-to-leading order 
calculation~\cite{chargedHiggs}.  The superparticle contributions are 
mostly computed at one loop order.  To evaluate these contributions, 
we also take into account corrections in powers of $\tan \beta$, which 
are important for large $\tan\beta$~\cite{Degrassi:2000qf}.  
The calculated branching ratio should be compared with the recent 
measurement ${\rm Br}(b \to s \gamma)=3.41(0.36) \times 10^{-4}$ 
by Ref.~\cite{Chen:2001fj}.  Here we take a rather conservative range
\begin{equation}
  2.0 \times 10^{-4} < {\rm Br}(b \to s \gamma) < 4.5 \times 10^{-4}\ .
\end{equation}

Finally, we impose the experimental bounds on the superparticles masses.  
The models which predict the large SUSY contribution to the muon $g-2$ 
may contain some light superparticles.   Thus negative searches 
of superparticle set lower bounds of the masses of superparticles
and put constraints on the model.  In our analysis we require that 
all charged superparticles are heavier than $100\ {\rm GeV}$.  

\begin{figure}
    \begin{center}
        \scalebox{0.4}[0.4]{\includegraphics{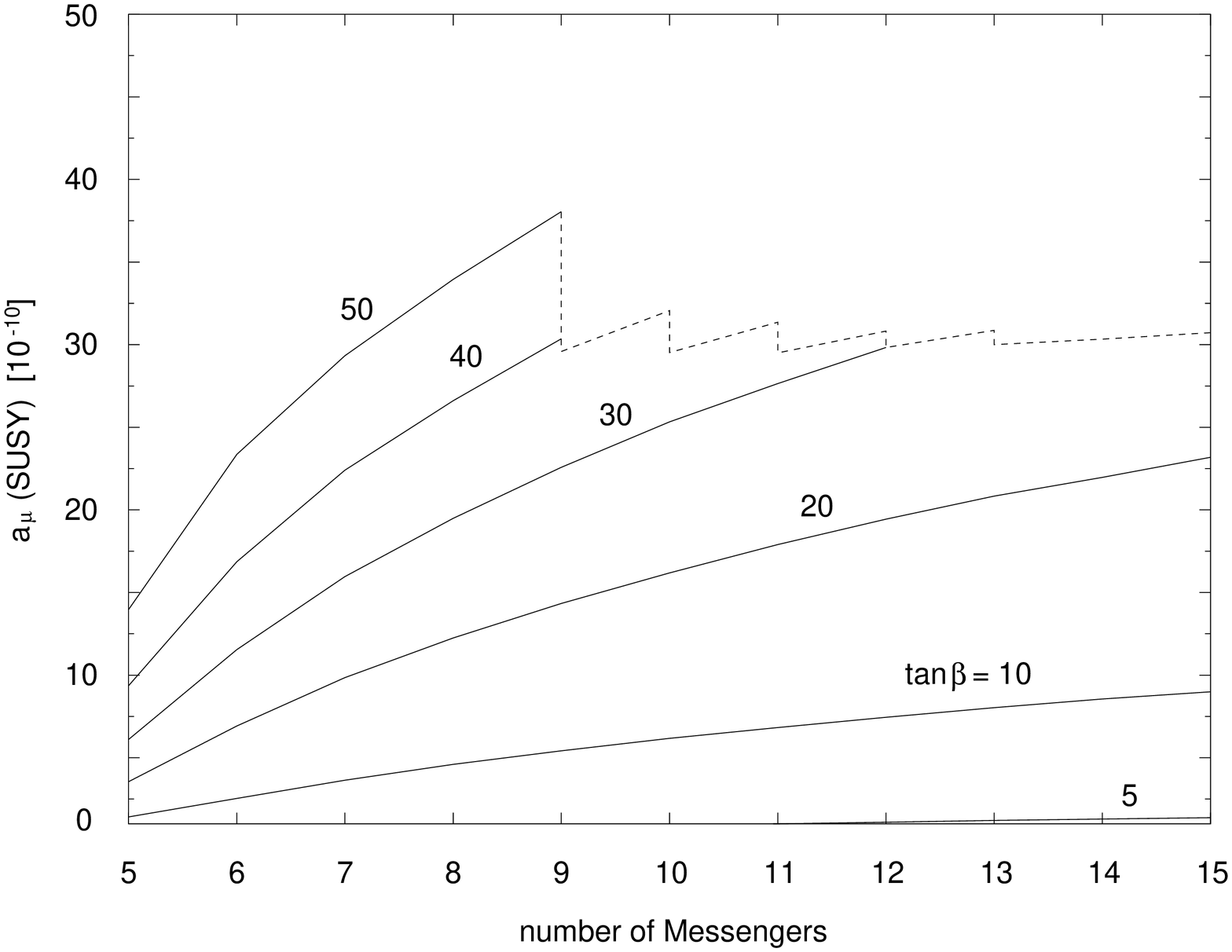}}
        \caption{The SUSY contribution to the muon $g-2$ 
          ($a_\mu({\rm SUSY})$) at the Higgs boson mass 
          $m_h = 114.1\ {\rm GeV}$ against the number of the pairs 
          of messengers in the deflected AMSB model.  
          We take $n = 3$ and $M_{\rm mess} = 10^{12}\ {\rm GeV}$.  
          The lines terminate by the negative search of the stau 
          for the large number of the messengers.}
        \label{fig:n3M12}
    \end{center}
    \begin{center}
        \scalebox{0.4}[0.4]{\includegraphics{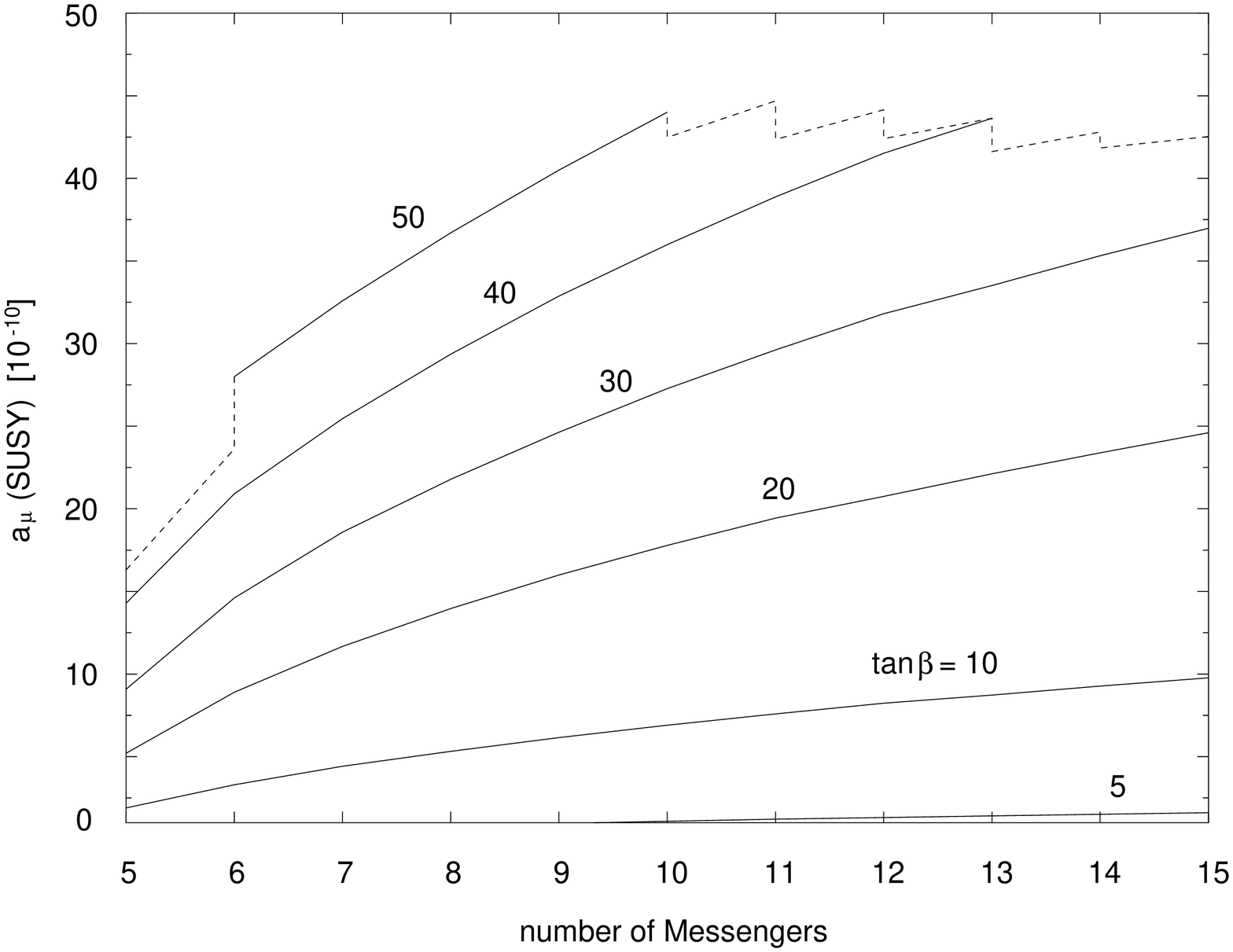}}
        \caption{Same as Fig.~\ref{fig:n3M12} but the messenger scale 
          $M_{\rm mess} = 10^{14}\ {\rm GeV}$.  
          The experimental bound from $b \to s \gamma$ excludes the 
          small number of the messengers for large $\tan\beta$, and 
          the lines terminate by the negative search of the stau 
          at the large number of the messengers.}
        \label{fig:n3M14}
    \end{center}
\end{figure}

\begin{figure}
    \begin{center}
        \scalebox{0.4}[0.4]{\includegraphics{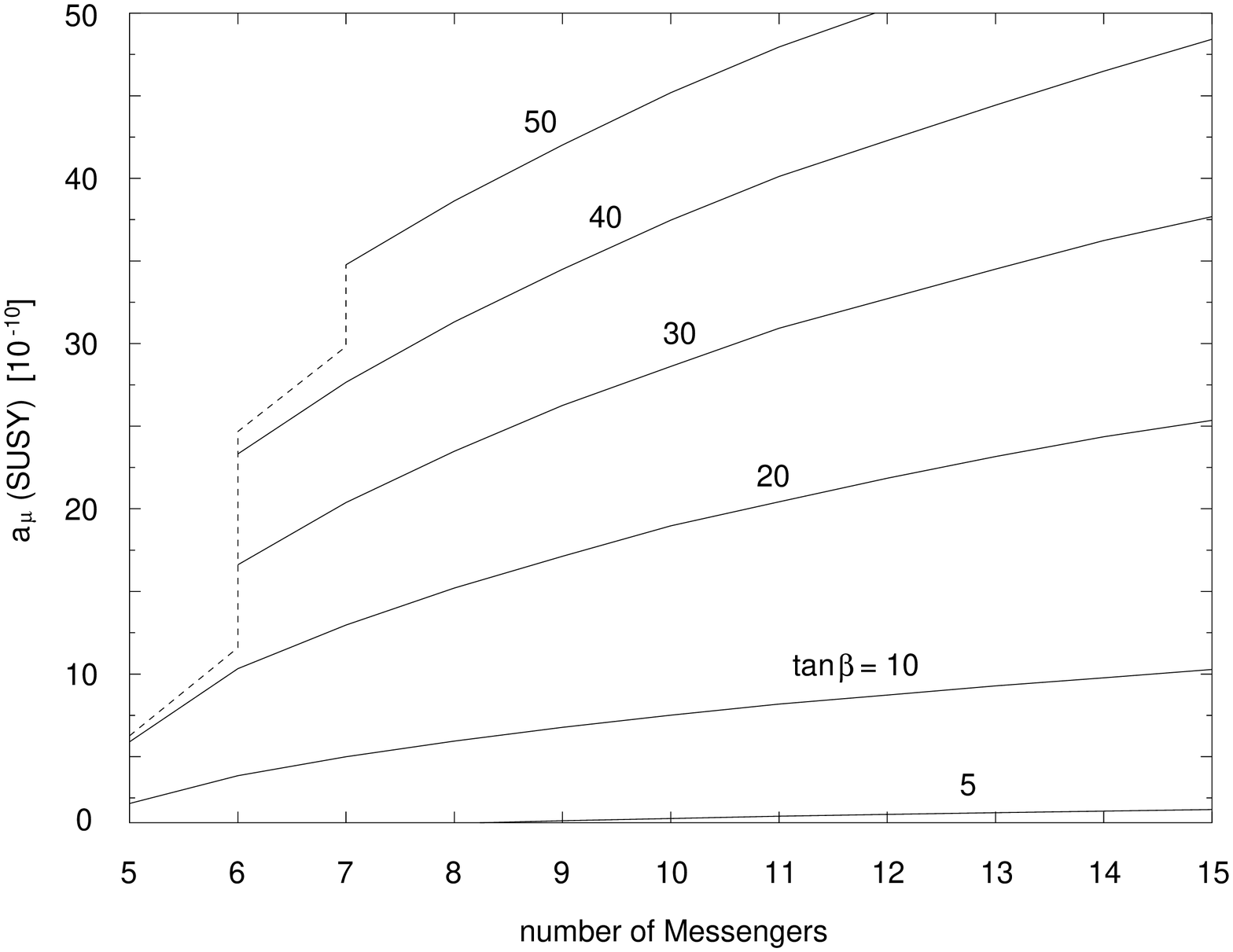}}
        \caption{Same as Fig.~\ref{fig:n3M12} but the messenger scale 
          $M_{\rm mess} = 10^{16}\ {\rm GeV}$.  
          The electroweak symmetry breaking does not occurs for the 
          small number of the messengers and especially 
          for large $\tan\beta$ the experimental bound from 
          $b \to s \gamma$ excludes the small number of the messengers.}
        \label{fig:n3M16}
    \end{center}
    \begin{center}
        \scalebox{0.4}[0.4]{\includegraphics{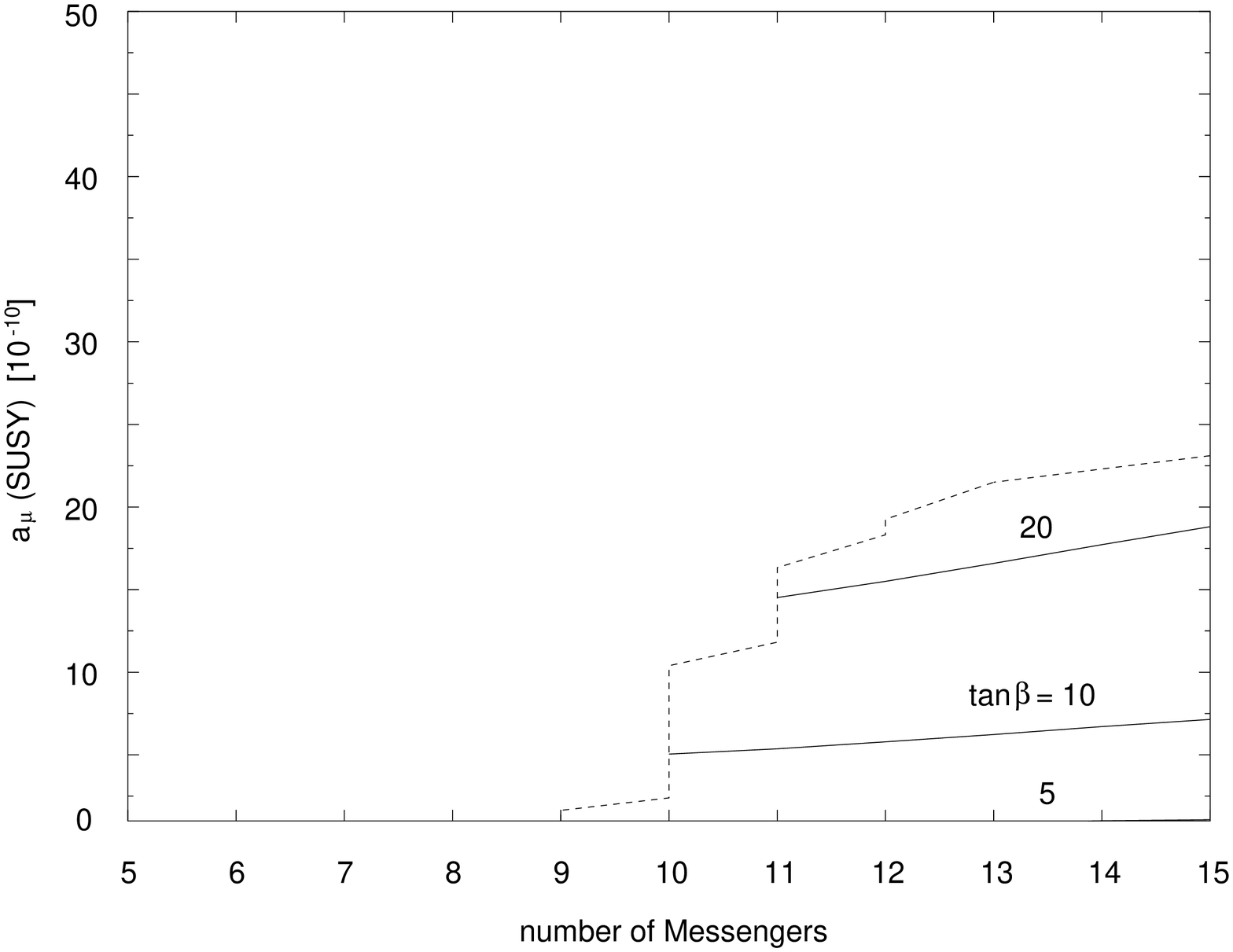}}
        \caption{The SUSY contribution to the muon $g-2$ 
          ($a_\mu({\rm SUSY})$) at the Higgs boson mass 
          $m_h = 114.1\ {\rm GeV}$ against the number of the pairs 
          of messengers in the deflected AMSB model.  
          We take $n = 4$ and $M_{\rm mess} = 10^{12}\ {\rm GeV}$.
          The electroweak symmetry breaking does not occurs for the 
          small number of the messengers and especially 
          for large $\tan\beta$ the experimental bound from 
          $b \to s \gamma$ excludes the small number of the messengers.  
          The negative search of the stau also excludes the region of 
          the large number of the messengers. }
        \label{fig:n4M12}
    \end{center}
\end{figure}

The results of numerical analysis are shown in 
Figs.~\ref{fig:n3M12}--\ref{fig:n3M16}.  These are the case of $n=3$.  
Here we determine $F_\Phi$ such that the Higgs boson mass becomes 
$m_h = 114.1\ {\rm GeV}$ and plot the values of $a_\mu({\rm SUSY})$ 
with several values of $\tan\beta$.  We consider $\tan\beta$ in 
the range $5 < \tan\beta < 50$~\footnote{
  Larger $\tan\beta$ actually makes the muon $g-2$ larger, but too large 
  $\tan\beta$ ({\it e.g.} $> 50$) makes the bottom Yukawa coupling blow up
  below the GUT scale.  
}.  
We also change $N_{\bf 5}$ and $M_{\rm mess}$.  
At first, we find that the SUSY contribution to the muon $g-2$ become 
larger as the number of the messengers increases.  This is because, 
for larger $N_{\bf 5}$, gaugino masses become larger, that is, $F_\Phi$ 
becomes smaller with the fixed value of Higgs mass.  Thus the SUSY 
contribution to the muon $g-2$ is enhanced when $N_{\bf 5}$ increases.  
Secondly, larger messenger scale $M_{\rm mess}$ makes the muon 
$g-2$ larger with $N_{\bf 5}$ fixed.  
This behavior is caused by the renormalization group effects, 
that is, if the messenger scale increases, the renormalization group 
effects make colored superparticles heavier compared to the 
uncolored ones.  Thus, the muon $g-2$ is enhanced as the messenger 
scale larger with the Higgs boson mass fixed.  
Finally, we can see that the deviation of the muon $g-2$ favors 
large $\tan\beta$.  This is because of the Higgs mass bound.  
From the results of the case of $n=3$, the muon $g-2$ reaches 
1$\sigma$ region of the deviation with $\tan\beta\gtrsim 20$.  

The contours of constant $\tan\beta$ terminate 
at the dotted lines~\footnote{
  Since the number of messengers is integer, the bounds become saw-like. 
  But we draw the constant $\tan\beta$ lines for convenience.  
} 
which show the bounds from the experimental constraints and 
the absence of the electroweak symmetry breaking.  
First, the negative search of the right-handed stau excludes 
the region of large $N_{\bf 5}$ and $\tan\beta$.  
This is because when $N_{\bf 5}$ becomes large, the sfermions becomes 
lighter (See Eq.~(\ref{m^2_f})) and larger $\tan\beta$ drive 
the sfermion masses smaller by the renormalization group effects.  
Secondly, we notice that the region of small $N_{\bf 5}$ is excluded.  
In this region with small $\tan\beta$, the electroweak symmetry 
breaking does not occur.  On the other hand, for large $\tan\beta$ 
the SUSY contribution to $b \to s \gamma$ becomes too large.  

\begin{figure}
    \begin{center}
        \scalebox{0.4}[0.4]{\includegraphics{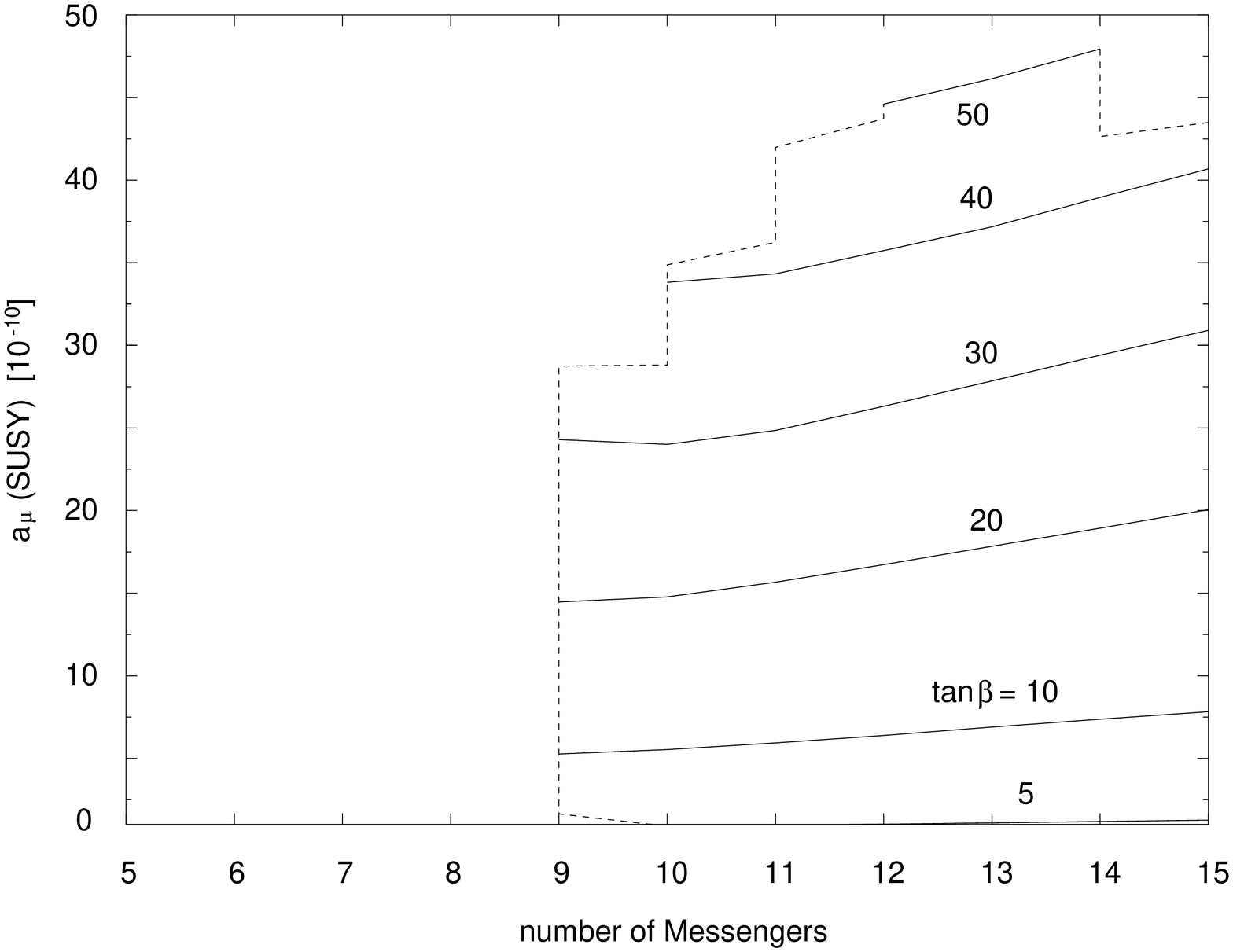}}
        \caption{Same as Fig.~\ref{fig:n4M12} but the messenger scale 
          $M_{\rm mess} = 10^{14}\ {\rm GeV}$.}
        \label{fig:n4M14}
    \end{center}
    \begin{center}
        \scalebox{0.4}[0.4]{\includegraphics{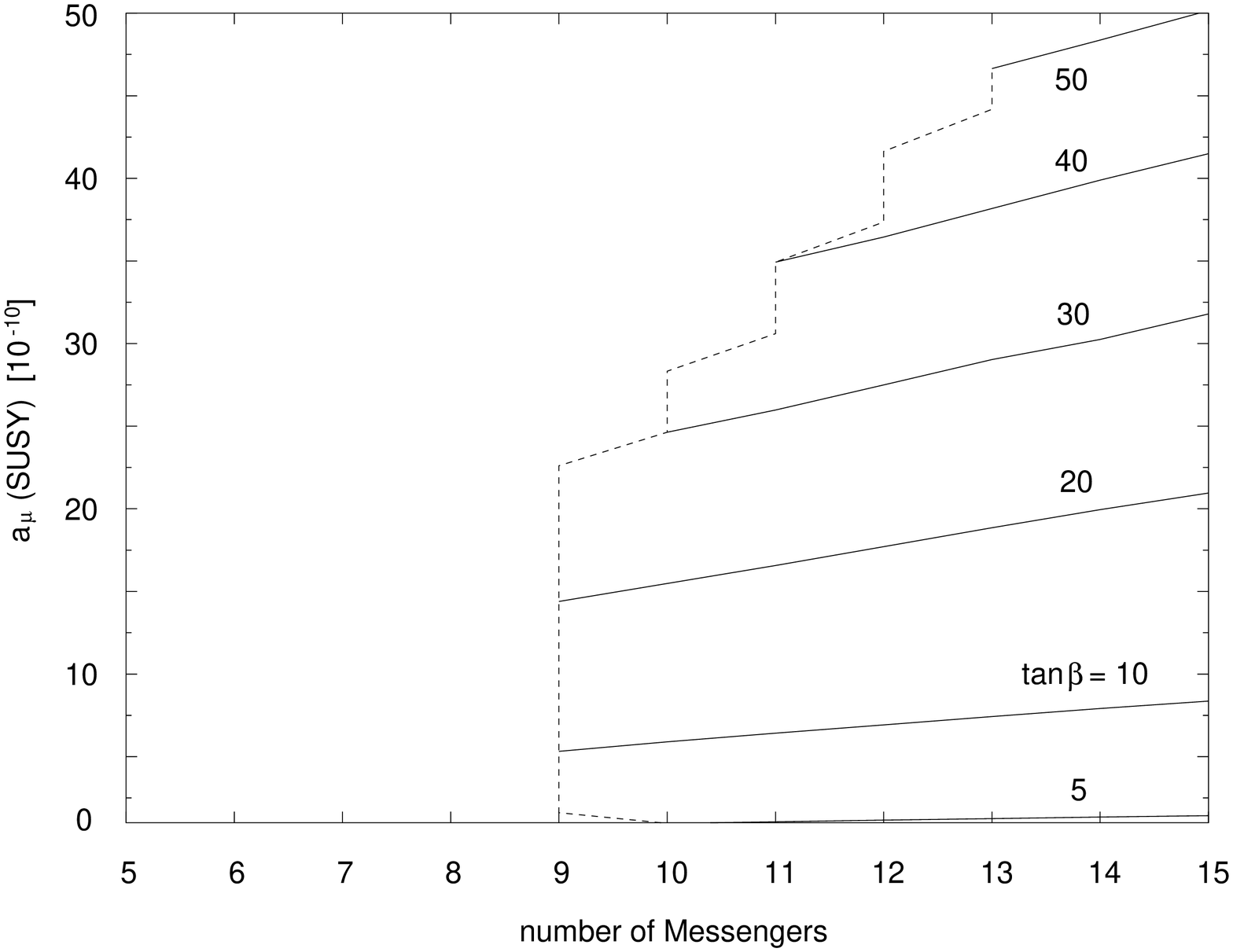}}
        \caption{Same as Fig.~\ref{fig:n4M12} but the messenger scale 
          $M_{\rm mess} = 10^{16}\ {\rm GeV}$.}
        \label{fig:n4M16}
    \end{center}
\end{figure}

In Figs.~\ref{fig:n4M12}--\ref{fig:n4M16}, we consider the case 
of $n=4$.  Here the parameters except for $n$ are the same as 
Figs.~\ref{fig:n3M12}--\ref{fig:n3M16}, respectively.  
We can understand these results by noting that $n$ appears as the 
combination $N_{\bf 5}/(n-1)$ in the gaugino masses.  
That is, the results of $n=4$ largely resemble the results with smaller 
$N_{\bf 5}$ in the case of $n=3$.  On the other hand, the sfermion masses 
depend on $n$ and $N_{\bf 5}$ in the different way and they become smaller 
as $n$ increases with gaugino masses fixed.  This effect cannot be neglected 
for the uncolored superparticles.  Thus, the muon $g-2$ becomes larger 
with the Higgs boson mass almost fixed.  As for ${\rm Br}(b \to s \gamma)$, 
the constraint becomes severer because the mass hierarchy between 
colored and uncolored superparticles becomes larger.  The large $F_\Phi$ 
regions also become to be excluded by the absence of the electroweak 
symmetry breaking.

The phenomenology of the deflected AMSB model depends sensitively on 
a choice of $(N_{\bf 5},n)$.  We can see that less messengers make 
almost all region be excluded by the tachyonic sleptons for fixed $n$.  
In fact for $n = 3$, more than 4 pairs of messengers are required.  
On the other hand, when $n$ is increased with $N_{\bf 5}$ fixed, 
gauginos become lighter and sleptons tend to be tachyonic. 
Thus larger $n$ requires larger number of messengers.  
However too many messengers make the gauge couplings blow up 
below the GUT scale~\cite{axion}.  
Thus, the cases of $n \geq 5$ are not attractive.  

Here we comment on the LSP.  If we take $n=4$, the fermionic part of 
the singlet field $X$ is no longer the LSP.  In fact, the mass of 
$\tilde{X}$ is of order of gravitino and is much heavier than the MSSM 
particles.  Thus we investigate the LSP in the MSSM sector.  
In the deflected AMSB model, uncolored superparticles tend to be lighter 
than colored ones.   We can see that the right-handed stau or one of 
the higgsino-like neutralinos becomes the lightest.  
In the almost all region of Figs.~\ref{fig:n4M12}--\ref{fig:n4M16}, 
we found that the LSP is the stau and thus this region is 
cosmologically not favored if the stau is stable~\cite{KudoYama}.  
However, we do not take this result seriously because 
there is a possibility that some light particle, 
say the fermionic partner of the axion, exists.  
If we require the neutralino to be the LSP, very small parameter regions 
are allowed only when $N_{\bf 5}=9$ (in  Figs.~\ref{fig:n4M14} and 
\ref{fig:n4M16}) or $N_{\bf 5}=10$ (in  Fig.~\ref{fig:n4M16}).  

We should note here that our results depend on top quark mass $m_t$.  
The radiative correction to the lightest Higgs boson mass is sensitively 
enhanced for larger value of $m_t$.  Therefore, the constraint from the 
Higgs mass becomes looser when $m_t$ is larger.  In the above analysis, 
we use $m_t = 174.3\ {\rm GeV}$.  On the other hand, the study in the 
case $m_t = 179.4\ {\rm GeV}$ gives the heavier Higgs mass about 2-3 GeV 
compared to the previous case of $m_t = 174.3\ {\rm GeV}$ at the same 
values of $M_{\rm mess}$ and $F_\Phi$.  Thus the model is viable on the 
wider region with the SUSY contribution to the muon $g-2$ large enough.  

As a discussion, we touch on the case of negative $n$.  In this case, 
the sign of the wino mass is the same as that of the gluino mass 
as the case of positive $n$.  Furthermore the slepton mass squareds 
become positive with less messengers.  The strongly interacting $SU(N_c)$ 
gauge sector makes $n$ negative~\cite{Okada}, where $n$ is given by 
$n=-2/(N_c-1)$.  As a distinctive feature, the bino and the wino are much 
lighter than the gluino as opposite to the positive $n$ case.  
Thus the mass hierarchy between colored and uncolored superparticles 
becomes large.  Hence smaller $\tan\beta$ is required to enhance 
the muon $g-2$ with the Higgs mass bound satisfied.  
Then the LSP almost becomes the bino-like neutralino.  

In this letter, we studied the deflected AMSB scenario in the light of 
the recent result of the muon $g-2$ from the Brookhaven E821 experiment.  
The recent result of the muon $g-2$ suggests the deviation from the SM 
prediction, though the uncertainties of the SM prediction, especially 
from the leading hadronic contribution, have not been settled.  
The results of the muon $g-2$ and $b \to s \gamma$ require 
$sign(M_2 \mu_H)$ and $sign(M_3 \mu_H)$ to be positive simultaneously.  
The deflected AMSB is the only known model in AMSB which provides this 
feature with avoiding additional CP violating phase naturally.  
By the detailed analysis, we found that in the deflected AMSB model,  
the SUSY contribution to the muon $g-2$ becomes as large as 
$\mathcal{O}(10^{-9})$, which is sufficient for the 3$\sigma$ deviation 
of the E821 result, with other experimental constraints satisfied.  
In particular, the Higgs mass puts a severe constraint on the model and 
large $\tan\beta$ is favored to enhance the muon $g-2$.  

\setcounter{equation}{0}
\section*{Acknowledgment}

The authors thank Takeo Moroi and Masahiro Yamaguchi for reading this 
manuscript carefully and useful comments.

\clearpage

\end{document}